\documentclass[final,times,pdflatex,twocolumn,showpacs,superscriptaddress,nofootinbib]{elsarticle}
\usepackage{lipsum}
\usepackage{xcolor}
\usepackage{etoolbox}
\usepackage{graphicx}
\usepackage{wrapfig}
\usepackage{amsmath}
\usepackage{amsfonts}
\usepackage{amssymb}
\usepackage{multirow}
\usepackage{slashed}
\usepackage{physics}
\usepackage{ulem}
\usepackage[colorlinks=true, linkcolor=blue, citecolor=blue, urlcolor=blue]{hyperref}

\DeclareRobustCommand{\Eq}[1]{Eq.~\eqref{eq:#1}}

\DeclareRobustCommand{\fig}[1]{Fig.~\ref{fig:#1}}

\DeclareRobustCommand{\refcite}[1]{Ref.~\cite{#1}}

\newcommand\bets{\begin{table*}}
\newcommand\eets[1]{\label{tb:#1}\end{table*}}

\journal{Physics Letters B}
\begin{document}

\title{Nonperturbative Collins-Soper Kernel from Chiral Quarks with Physical Masses }

%
\author[affiliation1]{Dennis Bollweg}
\author[affiliation3]{Xiang Gao}
\ead{gaox@anl.gov}
\author[affiliation1]{Swagato Mukherjee}
\author[affiliation3]{Yong Zhao}
\address[affiliation1]{Physics Department, Brookhaven National Laboratory, Upton, New York 11973, USA}
\address[affiliation3]{Physics Division, Argonne National Laboratory, Lemont, Illinois 60439, USA}

\date{\today}

\begin{abstract}
We present a lattice QCD calculation of the rapidity anomalous dimension of quark transverse-momentum-dependent distributions, i.e., the Collins-Soper (CS) kernel, up to transverse separations of about 1 fm. This unitary lattice calculation is conducted, for the first time, employing the chiral-symmetry-preserving domain wall fermion discretization and physical values of light and strange quark masses.
The CS kernel is extracted from the ratios of pion quasi-transverse-momentum-dependent wave functions (quasi-TMDWFs) at next-to-leading logarithmic perturbative accuracy. Also for the first time, we utilize the recently proposed Coulomb-gauge-fixed quasi-TMDWF correlator without a Wilson line. We observe significantly slower signal decay with increasing quark separations compared to the established gauge-invariant method with a staple-shaped Wilson line. This enables us to determine the CS kernel at large nonperturbative transverse separations and find its near-linear dependence on the latter. Our result is consistent with the recent lattice calculation using gauge-invariant quasi-TMDWFs, and agrees with various recent phenomenological parametrizations of experimental data.
%
%
\end{abstract}

\maketitle

%
\section{Introduction}\label{sec:intro}

The parton transverse-momentum-dependent distributions (TMDs) are crucial for a three-dimensional understanding of parton motions within a hadron, offering a more comprehensive view than traditional one-dimensional parton distribution functions (PDFs). It sheds light not only on the intrinsic motion of partons in transverse directions but also on the interplay between the transverse momentum of quarks and the spin of nucleons or quarks themselves. This comprehensive perspective is crucial for a deep understanding of the dynamic and complex nature of nucleons. 
The accurate characterization of TMDs is also critical for interpreting experimental data from high-energy collisions, particularly in relation to the transverse momentum distributions of electroweak and Higgs bosons~\cite{Collins:1984kg, Grewal:2020hoc}. They are fundamental to precision measurements, such as determining the mass and width of the W boson~\cite{Bozzi:2019vnl,CDF:2022hxs}. As high-energy physics experiments continue to advance, the measurement of TMDs will become increasingly important. The ongoing and future experiments at facilities such as the Large Hadron Collider~\cite{Kikola:2017hnp,Feng:2022inv} and Electron-Ion Collider~\cite{Boer:2011fh,Accardi:2012qut,AbdulKhalek:2022hcn,Abir:2023fpo} are expected to profoundly enrich our knowledge of TMDs. This will not only enhance our grasp of hadron structure and nucleon spin but also contribute significantly to the broader field of particle physics. 

Central to the study and practical application of TMDs is the Collins-Soper (CS) kernel, which is responsible for the (rapidity) scale evolution of TMDs~\cite{Collins:1981uk,Collins:1981va}, enabling the consistent interpretation of experimental data across different energy scales. It is instrumental in connecting theoretical predictions with experimental observations. The TMDs and CS kernel can be extracted through the global analysis of experimental data including the semi-inclusive deep inelastic scattering (SIDIS) and Drell-Yan processes~\cite{Davies:1984sp,Ladinsky:1993zn,Landry:2002ix,Konychev:2005iy,Sun:2014dqm,DAlesio:2014mrz,Bacchetta:2017gcc,Scimemi:2017etj,Bertone:2019nxa,Scimemi:2019cmh,Bacchetta:2019sam,Hautmann:2020cyp,Bury:2022czx,Bacchetta:2022awv,Moos:2023yfa,Isaacson:2023iui,Aslan:2024nqg}. However, the nonperturbative nature of Quantum Chromodynamics (QCD) at low transverse momenta 
necessitates certain parametrizations and introduces model dependence. As a result, there is an increasing interest for these intrinsically nonperturbative quantities to be calculated directly from first-principles lattice QCD.

Although direct simulation of TMDs on the Euclidean lattice is impractical, as TMDs are defined on the light-cone, it has been demonstrated that they can be accessed through quasi-TMDs within the framework of Large-Momentum Effective Theory (LaMET)~\cite{Ji:2013dva,Ji:2014gla,Ji:2020ect}. The quasi-TMDs involve the matrix elements of equal-time gauge-invariant (GI) operators:
\begin{align}\label{eq:OGI}
\begin{split}
    O^{\rm GI}_{\Gamma}(\mathbf{b};\eta)=\overline{\psi}(\frac{\mathbf{b}}{2})\Gamma W_\sqsupset(\frac{\mathbf{b}}{2},-\frac{\mathbf{b}}{2}, \eta)  \psi(-\frac{\mathbf{b}}{2}),
\end{split}
\end{align}
with 
$\mathbf{b}=(b_\perp, b_z)$ 
covering both longitudinal ($b_z$) and transverse ($b_\perp$) directions, linked by a staple-shaped Wilson line $W_\sqsupset$ whose length is characterized by $\eta$. In the large momentum and $\eta\rightarrow\infty$ limit, the quasi-TMDs can be related to the light-cone TMDs through the perturbative factorization~\cite{Ji:2014hxa,Ji:2018hvs,Ebert:2018gzl,Ebert:2019okf,Ji:2019sxk,Ebert:2019okf,Ji:2019ewn,Ebert:2020gxr,Vladimirov:2020ofp,Ji:2020jeb,Ji:2021znw,Ebert:2022fmh,Schindler:2022eva,Zhu:2022bja,Rodini:2022wic}. Building on this, significant advancements have been achieved over the past few years. The CS kernel has been extracted from either quasi-TMD parton distribution functions (TMDPDFs)~\cite{Shanahan:2020zxr,Shanahan:2021tst,Shu:2023cot}, quasi-TMD wave functions (TMDWFs)~\cite{LatticeParton:2020uhz,Li:2021wvl,LPC:2022ibr,Shu:2023cot,LatticePartonLPC:2023pdv,Avkhadiev:2023poz,Avkhadiev:2024mgd}, or the moments of the quasi-TMDs~\cite{Schlemmer:2021aij}.
Lattice QCD calculations of soft functions~\cite{LatticeParton:2020uhz,Li:2021wvl,LatticePartonLPC:2023pdv}, along with the first results of nucleon TMD PDFs~\cite{LPC:2022zci} and pion TMDWFs~\cite{Chu:2023jia} also have been reported. Additionally, progress has also been made in the systematical control of these calculations, including improved matching up to two loops~\cite{delRio:2023pse,Ji:2023pba}, addressing the operator mixing and working at physical quark masses~\cite{Ji:2017oey,Constantinou:2019vyb,Shanahan:2019zcq,Green:2020xco,Ji:2021uvr,Alexandrou:2023ucc,Avkhadiev:2023poz,Spanoudes:2024kpb,Avkhadiev:2024mgd}.

Despite notable progress, the lattice calculation of TMDs remains challenging. To suppress power corrections, a large momentum $P_z$ is required, which incurs a significant computational cost. In addition, the signal-to-noise ratio of the quasi-TMD matrix elements is adversely affected by exponential decay as the total length of the space-like Wilson line increases. This decay makes it particularly difficult to investigate quasi-TMDs at large $b_\perp$, where the results are desired to complement the phenomenological analysis.  What's more, the linear divergence and pinch-pole singularity in the Wilson lines also complicate the renormalization procedure, although they could be cancelled by the Wilson loop~\cite{Zhang:2022xuw} or held fixed by keeping a constant length of the Wilson line for given $b_\perp$~\cite{Avkhadiev:2023poz}. Besides, the operator mixings of Wilson-line operators~\cite{Constantinou:2019vyb,Shanahan:2019zcq,Green:2020xco,Ji:2021uvr,Alexandrou:2023ucc} also need to be subtracted systematically~\cite{Avkhadiev:2023poz}. Recently, a novel approach has been proposed for computing parton physics in the Coulomb gauge (CG)\cite{Gao:2023lny}, notably without the use of Wilson lines. Thereby, the complexity induced by the Wilson line can be avoided. It has been demonstrated that, in the large momentum limit, the CG quasi-PDF falls into the same universality class as the GI case under the LaMET framework. Further progress has been made in the realm of quasi-TMDs~\cite{Zhao:2023ptv}, involving the matrix elements of equal-time operators,
\begin{align}\label{eq:OCG}
\begin{split}
    O^{\rm CG}_{\Gamma}(\mathbf{b})=\overline{\psi}(\frac{\mathbf{b}}{2})\Gamma  \psi(-\frac{\mathbf{b}}{2})|_{\nabla \cdot \textbf{A}=0},
\end{split}
\end{align}
with the CG condition $\nabla \cdot \textbf{A}=0$ but without a Wilson line. The factorization of quasi-TMDs in the CG has been derived from the soft collinear effective theory (SCET)~\cite{Zhao:2023ptv} and verified at one loop in perturbation theory~\cite{Zhao:2023ptv,Liu:2023onm}. In this study, we have, for the first time, computed the quasi-TMDWFs of the pion in the CG and extracted the CS kernel from these measurements. 
Without Wilson lines, the CG correlators are multiplicatively renormalizable and free from the linear divergence~\cite{Gao:2023lny} and pinch singularity, as well as the operator mixings originating from the Wilson line geometry. Through our calculation, we show that, the CG approach leads to consistent CS kernel with the conventional GI approach. 
Moreover, the CG approach can significantly reduce the signal-to-noise ratio and extend the prediction power of lattice computation in the nonperturbative regime of interest to TMD physics.

%
\section{Theoretical framework}\label{sec:theo}

The pion quasi-TMDWF in the CG is defined as the Fourier transform of the matrix elements:
\begin{align}\label{eq:OCG}
	\tilde{\phi}_\Gamma^{\rm CG}(b_\perp,b_z,P_z,\mu) = \langle\Omega|O^{\rm CG}_{\Gamma}(\mathbf{b})|\pi^+;P_z\rangle,
\end{align}
where the pion is boosted with momentum $\textbf{P}=(0,0,P_z)$. By selecting $\Gamma=\gamma_t\gamma_5$ or $\gamma_z\gamma_5$, the quasi-TMDWF $\tilde{\phi}_\Gamma(x,b_\perp,P_z,\mu)$ can be related to the light-cone TMDWF $\phi(x,b_\perp,\zeta,\mu)$ (under the principle-value prescription of the light-cone Wilson lines~\cite{Belitsky:2002sm}) in the large $P_z$ limit through perturbative factorization, which can be expressed as~\cite{Ji:2019sxk,Ji:2021znw,Zhao:2023ptv},
\begin{align}\label{eq:qTMDWFfac}
\begin{split}
	&\frac{\tilde{\phi}_\Gamma(x,b_\perp,P_z,\mu)}{\sqrt{S_r(b_\perp,\mu)}}=H(x,\bar{x},P_z,\mu)\phi(x,b_\perp,\zeta,\mu)\\
	&\times\exp\left[\frac{1}{4}\left(\ln\frac{(2xP_z)^2}{\zeta}+\ln\frac{(2\bar{x}P_z)^2}{\zeta}\right)\gamma^{\overline{\rm MS}}(b_\perp,\mu)\right]\\
	&+\mathcal{O}\left(\frac{\Lambda_{\rm QCD}^2}{(xP_z)^2},\frac{1}{(b_\perp (xP_z))^2},\frac{\Lambda_{\rm QCD}^2}{(\bar{x}P_z)^2},\frac{1}{(b_\perp (\bar{x}P_z))^2}\right),
\end{split}
\end{align}
with $\bar{x}=1-x$. $\gamma^{\overline{\rm MS}}(b_\perp,\mu)$ is the CS kernel that governs the rapidity scale evolution from $\zeta$ to $(2xP_z)^2$ (or $(2\bar{x}P_z)^2$). $H(x,\bar{x},P_z,\mu)$ is a hard matching kernel that has been computed from one-loop perturbation theory~\cite{Zhao:2023ptv,Liu:2023onm}. $S_r(b_\perp,\mu)$ represents the reduced soft functions, extractable from the form factors of fast-moving color-charged states~\cite{Ji:2019sxk,Ji:2021znw,Zhao:2023ptv}. Consequently, the $x$-dependent light-cone TMDWF can be derived, subject to power corrections that are suppressed by large $P_z$ and $b_\perp$. Alternatively, the CS kernel can be extracted through the ratios of the quasi-TMDWFs with different momenta $P_1$ and $P_2$~\cite{Ji:2014hxa,Ebert:2018gzl,LatticeParton:2020uhz},
\begin{align}\label{eq:CSqTMDWF}
\begin{split}
	\gamma^{\overline{\rm MS}}(b_\perp,\mu)&=\frac{1}{\ln(P_2/P_1)}\ln\left[\frac{\tilde{\phi}(x,b_\perp,P_2,\mu)}{\tilde{\phi}(x,b_\perp,P_1,\mu)}\right]\\
	+&\delta\gamma^{\overline{\rm MS}}(x, \mu,P_1,P_2)+\rm p.c.,
\end{split}
\end{align}
with perturbative corrections $\delta\gamma^{\overline{\rm MS}}$ inferred from $H(x,\bar{x},P_z,\mu)$ and power corrections (p.c.).

\section{Lattice setup}\label{sec:setup}

The bare matrix elements of the pion quasi-TMDWF can be extracted from the two-point correlation functions in the lattice simulations. For CG quasi-TMDWFs, we compute,
\begin{align}\label{eq:CG2pt}
	C^{\rm CG}_{\pi {O}}(t_s;b_\perp,b_z,P_z) = \left \langle O^{\rm CG}_{\Gamma}(\mathbf{b},\mathbf{P}, t_s) \pi^\dagger(\textbf{y}_0,0)\right\rangle,
\end{align}
with,
\begin{align}
\begin{split}
	&O^{\rm CG}_{\Gamma}(\mathbf{b},\mathbf{P}, t_s) \\
	=&\sum_{\mathbf{y}} \overline{d}(\mathbf{y}+\frac{\mathbf{b}}{2}, t_s)\Gamma  u(\mathbf{y}-\frac{\mathbf{b}}{2}, t_s)|_{\nabla \cdot \textbf{A}=0} e^{-i \mathbf{P}\cdot (\mathbf{y}-\mathbf{y}_0)}.
\end{split}
\end{align}
Here $\textbf{y}_0$ is the source position, and $t_s$ is the time separation. We chose $\Gamma=\gamma_t\gamma_5$, as it should be free from the operator mixings caused by chiral symmetry breaking~\cite{Constantinou:2019vyb,Green:2020xco,Ji:2021uvr} under our lattice setup.

To improve the signal-to-noise ratio and increase the overlap with the pion ground state, we used extended pion source after boosted Gaussian smearing~\cite{Bali:2016lva},
\begin{align}
\pi^\dagger(\mathbf{y},t_s) = \overline{u}_s(\mathbf{y},t_s)\gamma_5 d_s(\mathbf{y},t_s),
\end{align}
with the $s$ denoting the smeared fields. We also compute the pion-pion two-point functions,
\begin{align}
	C_{\pi\pi}(t_s,P_z) = \left\langle \pi(\mathbf{P},t_s) \pi^\dagger(\mathbf{y}_0, 0) \right\rangle,
\end{align}
with smeared source and sink to extract the energy spectrum created by $\pi^\dagger$ as well as the overlap amplitudes. 

For the lattice simulation, we utilized a 2+1-flavor Domain-wall gauge ensemble generated by RBC and UKQCD Collaborations of size $N_s^3\times N_t\times N_5 =64^3\times 128\times12$, denoted by 64I~\cite{RBC:2023pvn}. The quark masses are at the physical point and the lattice spacing is $a^{-1}$ = 2.3549(49) GeV ($a=0.0836$ fm). For the boosted Gaussian smearing, the Gaussian radius was chosen to be $r_G$ = 0.58 fm, and we chose the quark boost parameter $j_z$ to be 0 and 6~\cite{Gao:2021xsm, Gao:2020ito} which are optimal to hadron momentum  $P_z=2\pi n_z/(N_sa)$ with $n_z$ = 0 and 8, so that the largest momentum in our calculation is $P_z=1.85$ GeV. Since only two-point functions are involved in this calculation, measurements at other momenta ($n_z\in[0,8]$) were also computed through contractions
using the same profiled quark propagator. 

To increase the statistics, we used 64 configurations coupled with All Mode Averaging (AMA) technique~\cite{Shintani:2014vja}. We computed 2 exact and 128 sloppy solutions for the quasi-TMDWFs with momenta $n_z\in[4,8]$, while 1 exact and 32 sloppy solutions for the cases with $n_z=[0,3]$. The quark propagators are evaluated from CG-fixed configurations using deflation based solver with 2000 eigen vectors.

After fixing the CG, the GI quasi-TMDWF defined from \Eq{OGI} shares the same quark propagators as the CG case but needs an additional staple-shaped Wilson line to maintain the gauge invariance. Therefore, we also computed the GI quasi-TMDWF correlators $C^{\rm GI}_{\pi {O}}(t_s;b_\perp,b_z,P_z,\eta)$ during the contraction. We chose $\eta=12a$ in this case using the same setup of the staple-shaped Wilson line as \refcite{Avkhadiev:2023poz}. 
We employed Wilson flow~\cite{Luscher:2010iy}, with a flow time $t_F=1.0$ (roughly smears the gauge fields over the radius $\sqrt{8a^2}$), to suppress the ultraviolet (UV) fluctuations and enhance the signal-to-noise ratio.

%
\section{Quasi-TMDWF}\label{sec:quasi}

The pion-pion and quasi-TMDWF correlators have the following spectral decompositions,
\begin{align}\label{eq:spectrumPP}
\begin{split}
	&C_{\pi\pi}(t_s;P_z) 	= \sum_{n=0}^{N_{\rm st}-1} \frac{|Z_n|^2}{2 E_n} \left (e^{-E_n t_s} + e^{-E_n(L_t-t_s)} \right),
\end{split}
\end{align}
and,
\begin{align}\label{eq:spectrumPO}
\begin{split}
	&C_{\pi O}^{\rm CG}(t_s;b_\perp,b_z,P_z) \\
	=&\sum_{n=0}^{N_{\rm st}-1} \frac{Z_n}{2 E_n} \langle\Omega|O_{\gamma_t\gamma_5}^{\rm CG}|n\rangle  (e^{-E_n t_s} + e^{-E_n(L_t-t_s)}),
\end{split}
\end{align}
where $E_n(P_z)$ is the energy level, and $Z_n = \langle n|\pi^\dagger(P_z)|{\Omega}\rangle$ is the overlap amplitude created by the pion interpolator (real and positive~\cite{Avkhadiev:2023poz}). $|{\Omega}\rangle$ represents the vacuum state, while $|n\rangle=|0\rangle,|1\rangle,...$ represents the ground state as well as the excited states.

\begin{figure}[th!]
    \centering
    \includegraphics[width=0.45\textwidth]{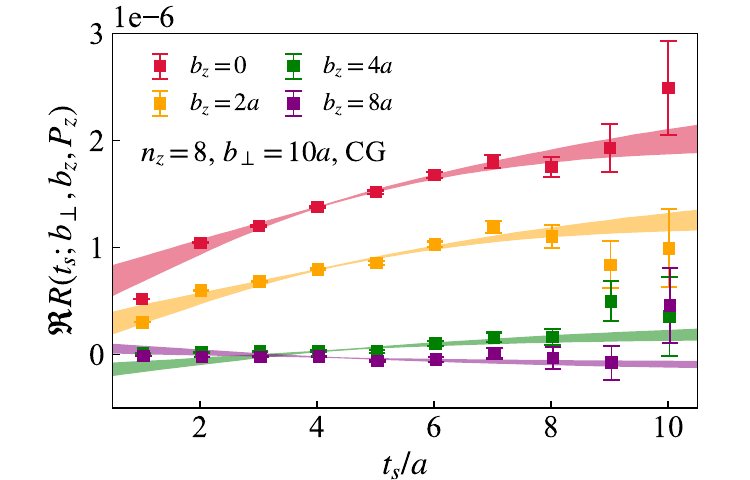}
	\caption{$R(t_s;b_\perp,b_z,P_z)$ as a function of $t_s$ for $n_z=8$ and $b_\perp=10a$. The bands are results from the two-state fits.\label{fig:2ptratio}}
\end{figure}

To take the advantage of high correlations between the pion-pion and quasi-TMDWF two-point functions, we construct their ratio as,
\begin{align}\label{eq:2ptratio}
\begin{split}
	R(t_s;b_\perp,b_z,P_z) &= \frac{-i C_{\pi O}^{\rm CG}(t_s;b_\perp,b_z,P_z)}{C_{\pi\pi}(t_s;P_z)}.
\end{split}
\end{align}
In \fig{2ptratio}, the ratios of our largest momentum $n_z=8$ at $b_\perp=10a$ are shown as an example. In the $t_s\rightarrow\infty$ limit, this ratio will reduce to $\langle\Omega|O_{\gamma_t\gamma_5}|0\rangle/ Z_0= E_0 \tilde{\phi}^B / Z_0$ and gives the bare quasi-TMDWF matrix elements $\tilde{\phi}^B(b_\perp,b_z,P_z,a)$. In practice, with finite $t_s$ we truncate \Eq{spectrumPP} and \Eq{spectrumPO} up to $N_{\rm st}=2$ and extract the bare matrix elements through the two-state fit. The fit results are shown as the bands in \fig{2ptratio} which can nicely describe the data point.

\begin{figure*}[th!]
    \centering\includegraphics[width=0.45\textwidth]{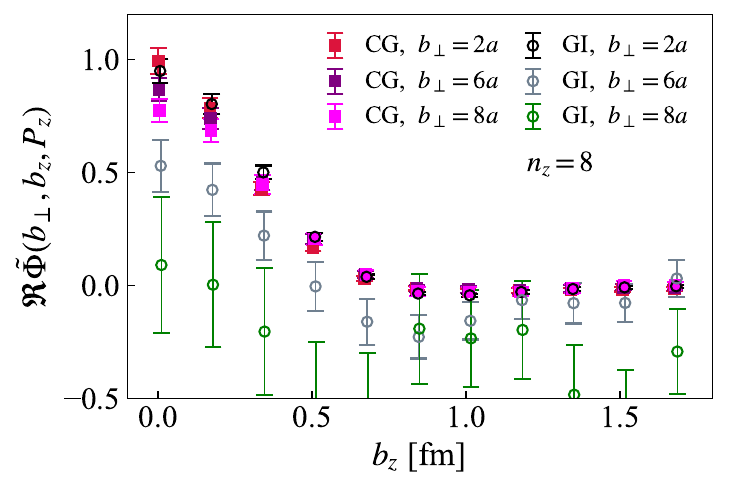}
    \includegraphics[width=0.45\textwidth]{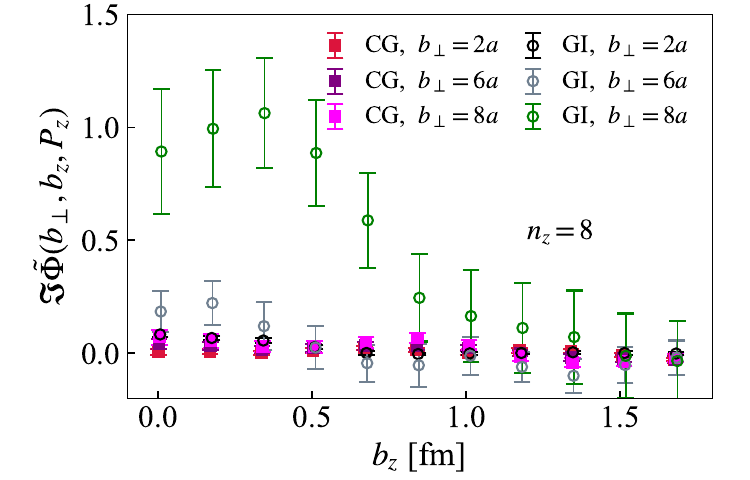}
	\caption{The real (left) and imaginary (right) parts of the renormalized quasi-TMDWF matrix elements at $n_z=8$ with $b_\perp=2a,6a,8a$ for the CG (filled squared symbols) and GI cases (open circled symbols).\label{fig:rbm}}
\end{figure*}

\begin{figure*}
    \centering
    \includegraphics[width=0.32\textwidth]{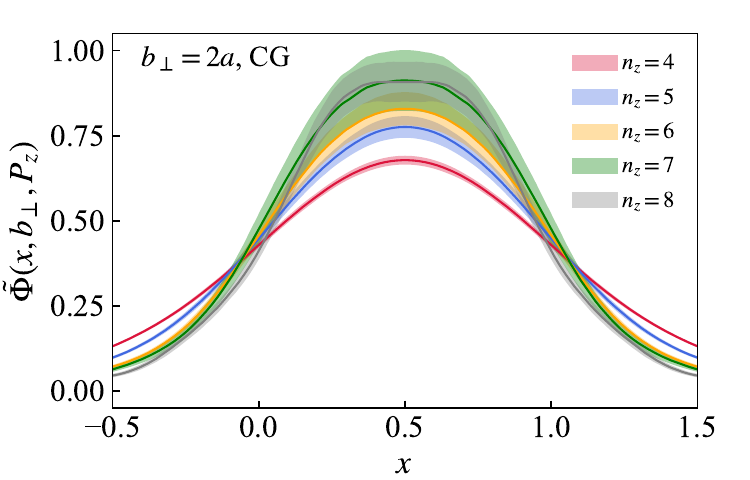}
    \includegraphics[width=0.32\textwidth]{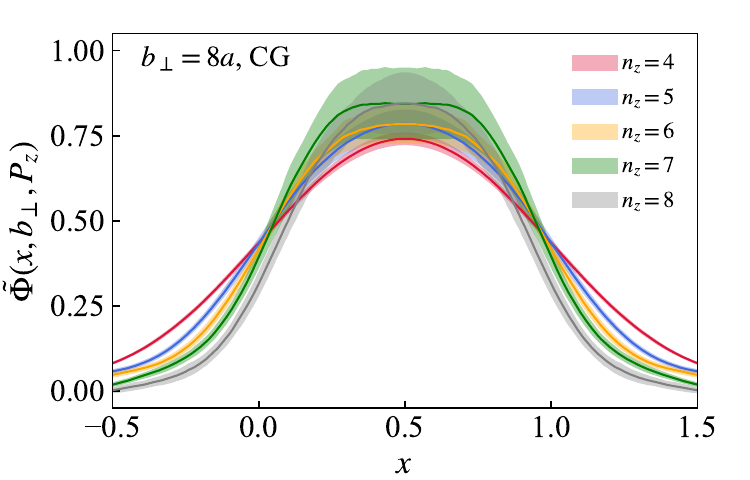}
    \includegraphics[width=0.32\textwidth]{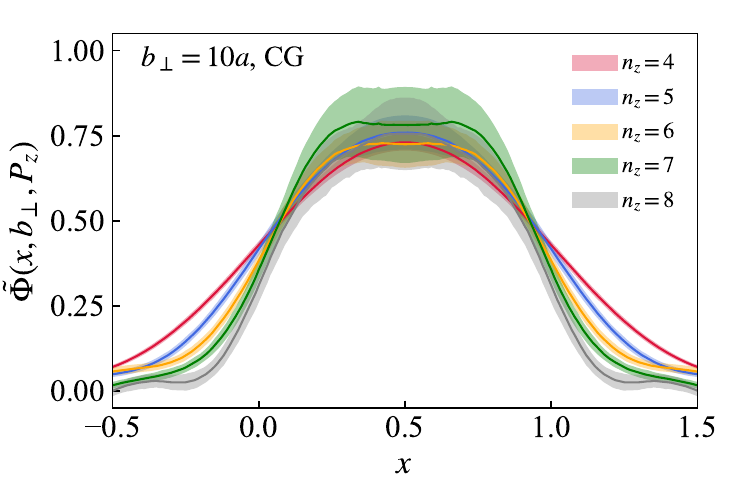}
    \includegraphics[width=0.32\textwidth]{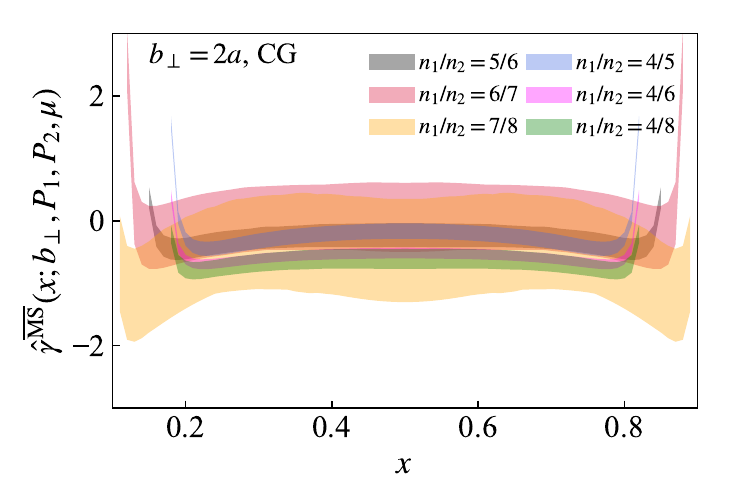}
    \includegraphics[width=0.32\textwidth]{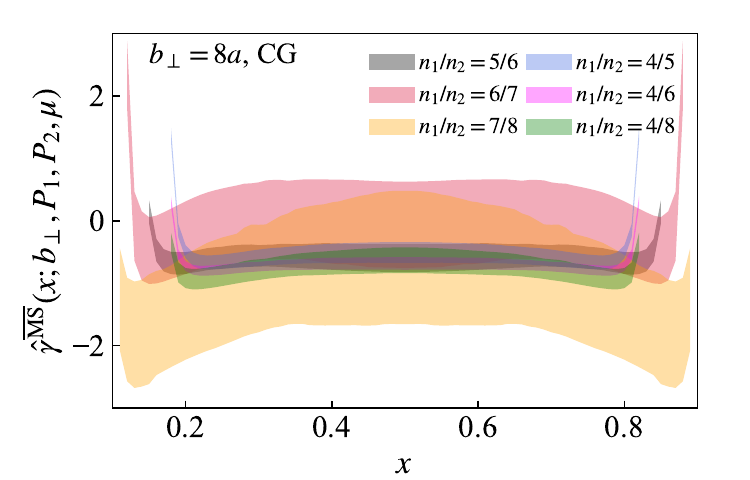}
    \includegraphics[width=0.32\textwidth]{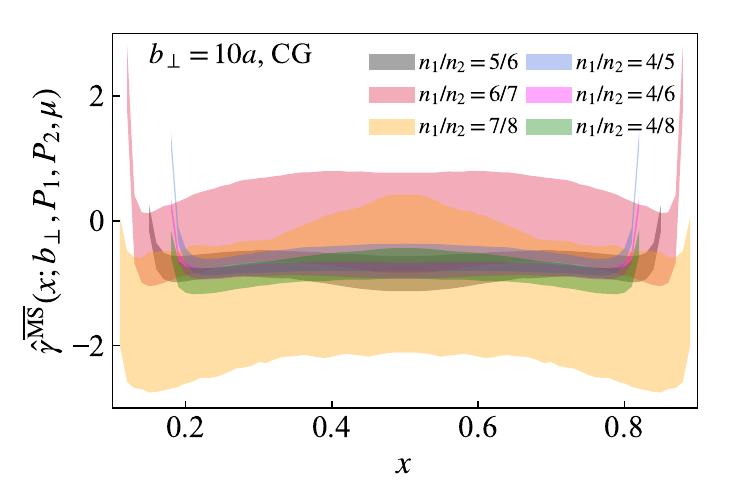}
	\caption{Upper panels: CG quasi-TMDWFs at momentum $n_z\in[4,8]$ and $b_\perp=2a,8a,10a$. Lower panels: CS kernel estimator $\hat{\gamma}^{\overline{\rm MS}}(x,b_\perp,P_1,P_2,\mu)$ derived from the ratio of the quasi-TMDWFs..\label{fig:CSx}}
\end{figure*}

The renormalization of CG quasi-TMD operator is straightforward. It involves only the CG quark wave function renormalization, which is an overall multiplicative constant and does not depend on the spatial separations $b_\perp$ and $b_z$~\cite{Gao:2023lny}.

In contrast, the GI quasi-TMD operator defined in \Eq{OGI}, though also subject to multiplicative renormalization, requires the removal of pinch pole singularities, cusp divergences, and linear divergences associated with the Wilson line~\cite{Ji:2017oey,Green:2017xeu,Ishikawa:2017faj,Zhang:2022xuw}. This renormalization is proportional to the total link length. 
In our implementation, the total length of the  staple-shape Wilson line is $2\eta+b_\perp$, independent of $b_z$.

Since the renormalization process solely involves the UV properties of operators and is independent of the external hadron states, 
we use the renormalization group invariant ratios~\cite{Avkhadiev:2023poz}
\begin{align}\label{eq:ratio}
\tilde{\Phi}(b_\perp,b_z,P_z) = \frac{\tilde{\phi}^B(b_\perp,b_z,P_z, \eta, a)}{\tilde{\phi}^B(b_\perp,0,0, \eta, a)},
\end{align}
without affecting the $x$- and $P_z-$dependences of the quasi-TMDWF after Fourier transform. The above ratio also may reduce some correlated uncertainties and eliminate some power corrections. Thus, we also adopt the same procedure for the CG matrix elements, whose renormalization is $b_\perp$ and $\eta$ independent.

In \fig{rbm}, we show the renormalized matrix elements for our largest momentum, $P_z=1.85$ GeV, and for $b_\perp=2a,6a$ and $8a$, as a function of $b_z$, for both CG (filled squared symbols) and GI (open circled symbols) cases. The left panel and right panel show the real and imaginary parts, respectively. It is evident that reasonable signal remains for the CG case even when $b_\perp$ become large. In contrast, the signal-to-noise ratio of the GI matrix elements rapidly deteriorates as $b_\perp$ increases, primarily due to the long Wilson line and its UV fluctuations. In addition, it is shown that the imaginary parts of the CG case are consistently zero, while they have non-zero values for GI case. This is expected as the imaginary part depends on the longitudinal orientation of the Wilson line in the GI case~\cite{Belitsky:2002sm}, whereas the CG condition does not favor any direction~\cite{Zhao:2023ptv}.

One can also observe that the matrix elements decrease as a function of $b_z$, diminishing to zero within the errors when $b_z\gtrsim1$ fm. This behavior facilitates the numerical Fourier transform to $x$-space with a simple truncation at the maximum value of $b_z$, which is expressed as,
\begin{align}
	\tilde{\Phi}(x,b_\perp,P_z)=\frac{P_z}{\pi}\int_0^{b_z^{\rm max}}e^{i(x-\frac{1}{2})P_zb_z}\tilde{\Phi}(b_\perp,b_z,P_z)
\end{align}
where we apply a first-order spline interpolation to smooth the data points. Since the CG quasi-TMDWF correlator is real and symmetric in $b^z$, the distribution must be real in the $x$-space.

In the upper panel of \fig{CSx}, we show selected results of the CG quasi-TMDWFs with momentum $n_z\in[4,8]$ and $b_\perp=2a,8a,10a$. Encouragingly, reasonable signal persists even when $b_\perp$ is as large as $10a$. However, the signal-to-noise ratio decreases as the momentum increases. In addition, it is evident that the quasi-TMDWFs, though appearing to be non-zero outside the physical region, have a trend to shrink into $x\in[0,1]$ as the momentum increases. This observation is consistent with the power expansion of the LaMET, suggesting the quasi-TMDWF is approaching the light-cone TMDWF in the large momentum limit.

%
\section{The Collins-Soper kernel}\label{sec:CS}

According to \Eq{CSqTMDWF}, we define the following estimator of the CS kernel utilizing the quasi-TMDWFs at finite momenta,
\begin{align}
\begin{split}
	\hat{\gamma}^{\overline{\rm MS}}(x,b_\perp,P_1,P_2,\mu)&=\frac{1}{\ln(P_2/P_1)}\ln\left[\frac{\tilde{\Phi}(x,b_\perp,P_2)}{\tilde{\Phi}(x,b_\perp,P_1)}\right]\\
	+&\delta\gamma^{\overline{\rm MS}}(x,\mu,P_1,P_2).
\end{split}
\end{align} 
In this work, we applied the perturbative corrections $\delta\gamma^{\overline{\rm MS}}$ derived from the next-to-leading logarithm (NLL) matching kernels for the CG case~\cite{Ji:2019ewn,Ebert:2022fmh,Avkhadiev:2023poz,Zhao:2023ptv} as only one-loop non-cusp anomalous dimension is available. The $\overline{\rm MS}$ scale has been set to be $\mu=2$ GeV. If the power corrections and higher-order corrections are small, $\hat{\gamma}^{\overline{\rm MS}}$ should be independent of $P_z$ and $x$.

In the lower panels of \fig{CSx}, we show the CS kernel estimators for the CG case with various combination of momenta, $n_1$ and $n_2$, as a function of $x$. 
The $x$-independent plateaus can be found in the moderate $x$ region within the errors, which is robust even at the largest $b_\perp$. This indicates the effectiveness of the factorization formula in \eqref{eq:qTMDWFfac}. In the end-point regions of both small and large $x$, the results appear to diverge, signaling a breakdown of the factorization in these areas. However, the length of plateaus extend as the momentum increases, which is consistent with the power corrections suggested in \Eq{qTMDWFfac}. 

As for the momentum dependence, it is absent for the case of large $b_\perp$, which indicates well suppressed power corrections by $1/(P_zb_\perp)$, despite their slightly larger errors. However, results at small $b_\perp$ (e.g., for $2a$) and with small momentum (e.g., for $n_1=4$) deviated from the ones derived from larger momenta. This momentum dependence is reduced when $n_1$ and $n_2$ gets close, and disappear when $n_1$ and $n_2$ are close enough (e.g., for $n_1/n_2=4/5$). This observation suggests that the power corrections and higher-order perturbative corrections are not well suppressed in the cases of small $b_\perp$ and large differences in $P_z$. 

To estimate the CS kernel, we averaged over the estimator $\hat{\gamma}^{\overline{\rm MS}}(x, b_\perp,\mu,P_1,P_2)$ within $x\in[x_0,1-x_0]$ across various $n_1$ and $n_2$. Only the cases of $n_2-n_1=1$ are considered. The value of $x_0$ is determined by requiring $2x_0P_zb_\perp>1$ and $2x_0P_z>0.7~{\rm GeV}$, suggested by the power correction. As a result, $b_\perp=a$ is always excluded in this work. The averages over $x$ and different valeus $n_1/n_2$ are carried out for each bootstrap sample of gauge configurations. The results are quoted from the median and 68\% confidence limit of the distribution of all bootstrap samples. Thus, our quoted errors include the correlated statistical and systematic errors arising from $x$ and $P_z$ averaging.

\begin{figure}[t!]
    \centering
    \includegraphics[width=0.48\textwidth]{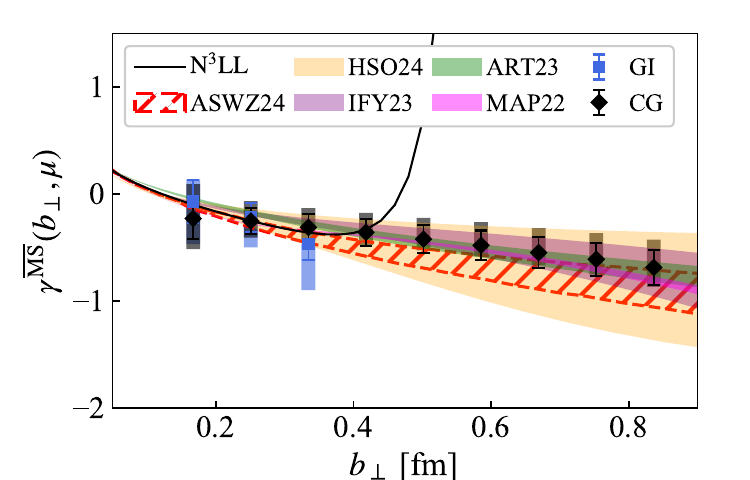}
	\caption{The CS kernel determined from the CG quasi-TMDWFs are shown as the black points and patches, which represent the exclusion and inclusion of momentum pairs $n_1/n_2=6/7$ and $7/8$, respectively. The averages including $n_1/n_2=6/7$ and $7/8$ are depicted as black patches under the data points. The results from GI quasi-TMDWFs calculated in this work are shown as the blue points and patches. For comparison, the CS kernels from recent phenomenological parameterizations of experimental data are shown from MAP22~\cite{Bacchetta:2022awv}, ART23~\cite{Moos:2023yfa}, IFY23~\cite{Isaacson:2023iui} and HSO24 (E605)~\cite{Gonzalez-Hernandez:2022ifv,Aslan:2024nqg}. We also show the perturbative results ($\rm N^3LL$) from \refcite{Vladimirov:2016dll, Li:2016ctv}, as well as a recent lattice calculation (ASWZ24) from GI quasi-TMDWFs in the continuum limit with high statistics~\cite{Avkhadiev:2024mgd}.\label{fig:CS}}
\end{figure}

Our results for the CS kernel are shown as the black points in \fig{CS}. 
The error bars indicate errors when $n_1/n_2=6/7$ and $7/8$ are excluded from the average. The averages including $n_1/n_2=6/7$ and $7/8$ are depicted as black patches under the data points. 

The CS kernel extracted from the GI quasi-TMDWFs calculated in this work is also shown as the blue points and patches, which is consistent with the CG case at smaller $b_\perp$. We do not show the CS kernel from the GI quasi-TMDWFs at $b_\perp >4a$ because the results are too noisy for comparison as already indicated in \fig{rbm}. It has been demonstrated in Ref.~\cite{Avkhadiev:2024mgd} that after the matching correction, which takes into account of the power corrections at small $b_\perp$~\cite{Avkhadiev:2023poz} and the so-called linear renormalon subtraction at large $b_\perp$~\cite{Liu:2023onm}, the imaginary part of the CS kernel in the GI case is consistent with zero, so we only take the real part of the final result. 

Our results agree with the $\rm N^3LL$ perturbative prediction~\cite{Vladimirov:2016dll, Li:2016ctv} at the short distances ($b_\perp\lesssim$ 0.4 fm).
Beyond this point, the perturbative prediction becomes sensitive to  the Landau pole and, thereby, loses reliability.

Although our CS kernel from the GI case loses signal for $b_\perp\gtrsim0.4$~fm, our CG results continue to show very good signals at $b_\perp$ up to about $1$~fm. 

For comparisons, we show the most recent lattice QCD calculation (ASWZ24) from GI quasi-TMDWFs in the continuum limit with high statistics~\cite{Avkhadiev:2024mgd}. Evidently, the results from CG and GI case are consistent with each other, suggesting they fall into the same universality class in the large $P_z$ limit under the framework of LaMET~\cite{Gao:2023lny}.

Furthermore, our nonperturbative theoretical predictions of the CS kernel are in agreement with the recent phenomenological parameterizations of experimental data, MAP22~\cite{Bacchetta:2022awv}, ART23~\cite{Moos:2023yfa}, IFY23~\cite{Isaacson:2023iui}, and HSO24 (E605)~\cite{Gonzalez-Hernandez:2022ifv,Aslan:2024nqg}, which shows a near-linear $b_\perp$ dependence as proposed in Ref.~\cite{Collins:2014jpa}.

%
\section{Conclusion}\label{sec:conclusion}

We conducted the first lattice QCD calculation of the CS kernel utilizing the recently proposed CG quasi-TMD approach as well as employing unitary domain wall fermion discretization with physical quark masses and a fine lattice spacing. 

The CG approach shows significantly lower signal decay, allowing the CS kernel to be determined for extended transverse separations. At the same time, we show that our results are well compatible with the widely used gauge-invariant method. Notably, our results agree with the recent phenomenological parameterizations of experimental data and suggest a near-linear dependence of the CS kernel on large $b_\perp$. 

This work lays a solid foundation for future research into the CS kernel at larger values of $b_\perp$, and advances the QCD computations in the nonperturbative regime of TMD physics.


\section*{Acknowledgements}

Our calculations were performed using the Grid \cite{Boyle:2016lbp,Yamaguchi:2022feu} and GPT \cite{GPT} software packages. We thank Christoph Lehner for his advice on using GPT. We thank Rui Zhang and Jinchen He for valuable discussions, and Artur Avkhadiev for helpful comments on the manuscript. We also thank Ted Rogers and J. Osvaldo Gonzalez for communications on the HSO paramterization.

This material is based upon work supported by The U.S. Department of Energy, Office of Science, Office of Nuclear Physics through Contract No.~DE-SC0012704, Contract No.~DE-AC02-06CH11357, and within the frameworks of Scientific Discovery through Advanced Computing (SciDAC) award Fundamental Nuclear Physics at the Exascale and Beyond and the Topical Collaboration in Nuclear Theory 3D quark-gluon structure of hadrons: mass, spin, and tomography.
YZ is partially supported by the 2023 Physical Sciences and Engineering (PSE) Early Investigator Named Award program at Argonne National Laboratory.

This research used awards of computer time provided by: The INCITE program at Argonne Leadership Computing Facility, a DOE Office of Science User Facility operated under Contract DE-AC02-06CH11357; the ALCC program at the Oak Ridge Leadership Computing Facility, which is a DOE Office of Science User Facility supported under Contract DE-AC05-00OR22725; the National Energy Research
Scientific Computing Center, a DOE Office of Science User Facility
supported by the Office of Science of the U.S. Department of Energy
under Contract DE-AC02-05CH11231 using NERSC award
NP-ERCAP0028137. Part of the data analysis is carried out on Swing, a high-performance computing cluster operated by the Laboratory Computing Resource Center at Argonne National Laboratory.

\bibliographystyle{elsarticle-num}
\bibliography{ref}

\end{document}